\documentclass[aps,prb,twocolumn,superscriptaddress]{revtex4-1}
\usepackage{amsmath}
\usepackage{xcolor}
\usepackage{graphicx}
\usepackage{hyperref}

\begin{document}

\title{Tuning the two-step melting of magnetic orders in a dipolar kagome spin ice by quantum fluctuations}

\date{\today}

\author{Yao Wang}
\thanks{These two authors contributed equally.}
\affiliation{Institute of Physics, Chinese Academy of Sciences, Beijing 100190, China}
\affiliation{University of Chinese Academy of Sciences, Beijing 100049, China}
\author{Stephan Humeniuk}
\thanks{These two authors contributed equally.}
\affiliation{Institute of Physics, Chinese Academy of Sciences, Beijing 100190, China}
\author{Yuan Wan}
\email{yuan.wan@iphy.ac.cn}
\affiliation{Institute of Physics, Chinese Academy of Sciences, Beijing 100190, China}
\affiliation{University of Chinese Academy of Sciences, Beijing 100049, China}
\affiliation{Songshan Lake Materials Laboratory, Dongguan, Guangdong 523808, China}

\begin{abstract}
Complex magnetic orders in frustrated magnets may exhibit rich melting processes when the magnet is heated toward the paramagnetic phase. We show that one may tune such melting processes by quantum fluctuations. We consider a kagome lattice dipolar Ising model subject to transverse field and focus on the thermal transitions out of its magnetic ground state, which features a $\sqrt{3}\times\sqrt{3}$ magnetic unit cell. Our quantum Monte Carlo (QMC) simulations suggest that, at weak transverse field, the $\sqrt{3}\times\sqrt{3}$ phase melts by way of an intermediate magnetic charge ordered phase where the lattice translation symmetry is restored while the time reversal symmetry remains broken. By contrast, at stronger transverse field, QMC simulations suggest the $\sqrt{3}\times\sqrt{3}$ order melts through a floating Kosterlitz-Thouless phase. The two distinct melting processes are separated by either a multicritical point or a short line of first order phase transition. 
\end{abstract}

\maketitle

\section{Introduction \label{sec:intro}}

The notion of symmetry is fundamental to our understanding of magnetic orders. In a simple magnetic system such as Ising ferromagnet, the low temperature ferromagnetic phase spontaneously breaks the time reversal symmetry, which is captured by a single order parameter, namely the uniform magnetization. As the temperature increases, the ferromagnetic phase melts through a single thermal phase transition, whose universality is essentially fixed by the symmetry transformation properties of the order parameter and the spatial dimensionality.

The melting process of magnetic phases can be significantly richer in geometrically frustrated magnets, where frustration effects can result in complex magnetic ground states that spontaneously break multiple symmetries, corresponding to the development of several distinct but intertwined order parameters. As the system is heated toward the paramagnetic phase, the low temperature magnetic phase may melt in multiple steps through intermediate phases where the symmetries are partially restored. Depending on the specific contexts, the same low temperature magnetic phase may feature different multistep melting processes through different intermediate phases. The universality class of the melting transition then depends on the specific melting pathway as oppose to being fixed by the low temperature magnetic phase alone.

\begin{figure*}
\includegraphics[width = 0.75\textwidth]{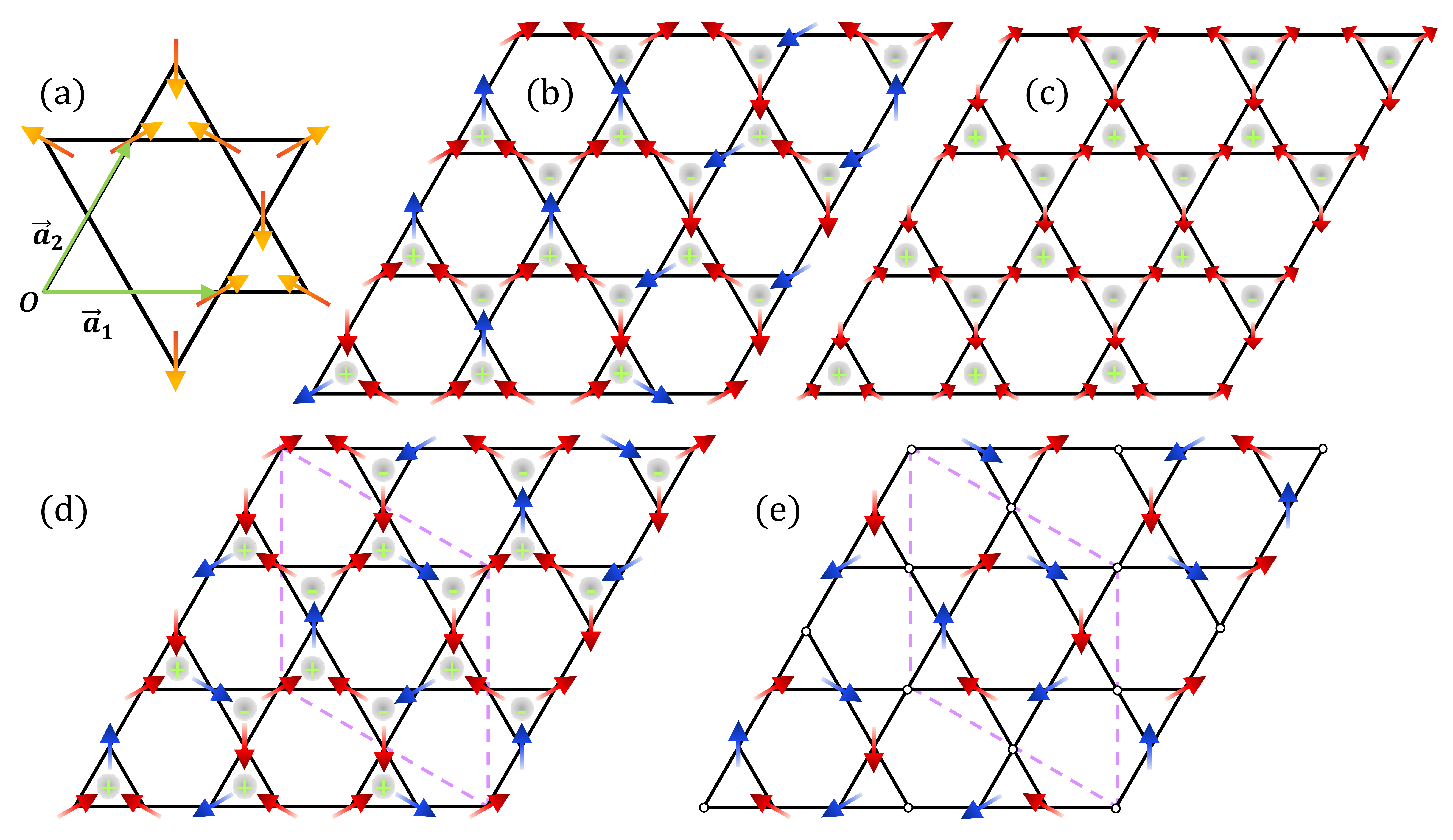}
\caption{(a) Kagome lattice. The green arrows are the primitive vectors $\mathbf{a}_{1,2}$ emanating from the origin $O$. The yellow arrow at each site shows the direction of the local spin $\hat{z}_i$ axis. (b) A snap shot of the spin configurations in the magnetic charge ordered phase. The arrows show the direction of the spins. Spins with $\sigma^{z}_i=1$ and $-1$ are respectively colored in red and blue. Here, the magnetic charges exhibit an ordered pattern --- the up triangles carry positive magnetic charge ($+$), whereas the down triangles carry negative magnetic charge ($-$). The spins, on the other hand, are fluctuating. (c) Magnetic moment in the magnetic charge ordered phase obtained by averaging over many snap shots similar to (b). As the up triangles have two spins pointing inward and one outward, the spins on average point toward the center of the up triangle. The size of magnetic moment is about $1/3$ of the length of the spin. (d) One of the six degenerate domains of the $\sqrt{3}\times\sqrt{3}$ phase. The spins are all ordered. Its magnetic charge distribution shows the same ordered pattern as (b). The purple dashed rhombus demarcates the magnetic unit cell. (e) The partially disordered phase. Open circles correspond to spins that fluctuate between $\sigma^z_i = 1$ and $-1$, showing no net magnetic moment. Note the partially disordered phase is \emph{not} observed in our model Eq.~\eqref{eq:hamiltonian}. 
\label{fig:cartoon}}
\end{figure*}

A prominent example where the multistep melting occurs is the two-dimensional dipolar kagome spin ice. In its simplest setting~\cite{Moeller2009,Chern2011,Rougemaille2011,Chern2012}, Ising spins form a kagome lattice with their magnetic moments lying \emph{in} the kagome plane. The spins interact through the long range magnetic dipole-dipole interaction. This system hosts at low temperature a magnetic ground state with $\sqrt{3}\times\sqrt{3}$ magnetic unit cell (Fig.~\ref{fig:cartoon}d), corresponding to a magnetic ordering wave vector $\mathbf{Q} = 2\mathbf{K}$, where $\mathbf{K}$ is at the K point of the first Brillouin zone. The $\sqrt{3}\times\sqrt{3}$ magnetic ground state breaks both the lattice translation symmetry and the time reversal symmetry. Upon heating, the $\sqrt{3}\times\sqrt{3}$ phase melts through an intermediate magnetic charge ordered phase where the lattice translation symmetry is restored but the time reversal symmetry remains broken. There, the magnetic charges inside the triangles of the kagome lattice exhibit an ordered pattern while the spins are fluctuating (Fig.~\ref{fig:cartoon} b\&{}c). In the example shown in Fig.~\ref{fig:cartoon}c, each up triangle carries one unit of positive magnetic charge, while each down triangle carries one unit of negative magnetic charge. This pattern respects the lattice translation symmetry. Equivalently, the static magnetic moments form a staggered, all-in-all-out pattern, which respects the lattice translation symmetry as well. The melting transition from the $\sqrt{3}\times\sqrt{3}$ phase to the magnetic charge ordered phase is of the three-state Potts universality. Further increasing the temperature finally brings the system to the paramagnetic phase through an Ising transition that restores the time-reversal symmetry.

The $\sqrt{3}\times\sqrt{3}$ phase exhibits a very different melting process in a closely related system. In the kagome Ising antiferromagnet with both nearest and second neighbor interactions~\cite{Kano1953,Wolf1988,Takagi1993,Wills2002}, the intermediate magnetic charge ordered phase is evaded. Instead, the $\sqrt{3}\times\sqrt{3}$ phase melts through a floating Kosterlitz-Thouless (KT) phase by two consecutive KT transitions~\cite{Chern2012}.

\begin{figure}
\includegraphics[width = \columnwidth]{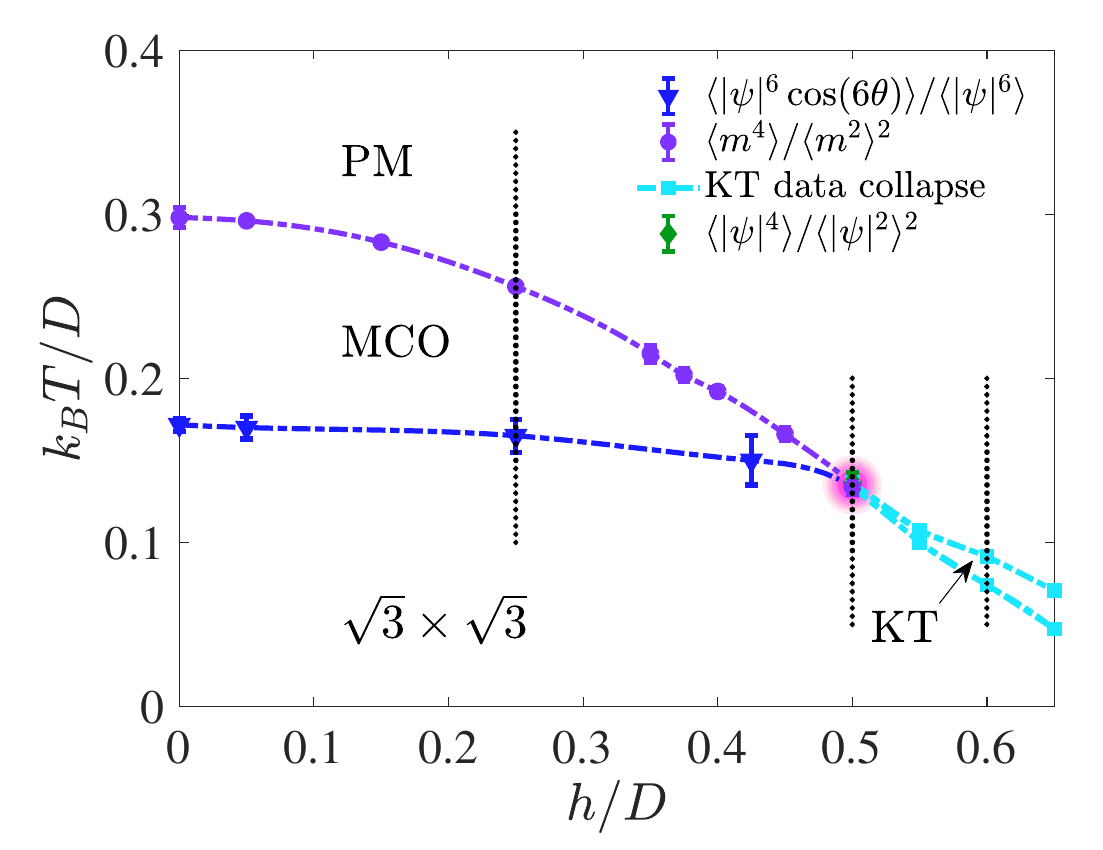}
\caption{Phase diagram of Eq.~\eqref{eq:hamiltonian} as a function of temperature $T$ and transverse field $h$, deduced from QMC simulations. We identify four phases: the paramagnetic phase (PM), the magnetic charge ordered phase (MCO), the Kosterlitz-Thouless phase (KT), and the $\sqrt{3}\times\sqrt{3}$ phase ($\sqrt{3}\times\sqrt{3}$). Different colors/symbols indicate the different methods employed to determine the phase boundaries (see the main text for detail). A possible multicritical point or a short line of first order transition may exist in the area shaded in magenta. We expect a quantum critical point at $T=0$, $h/D>0.65$, which is not determined in this work. Dashed lines mark the three temperature scans that shall be discussed in Sec.~\ref{sec:results}.
\label{fig:phase_diag}}
\end{figure}

The rich melting processes of the $\sqrt{3}\times\sqrt{3}$ phase in the dipolar kagome spin ice and related systems make one wonder if there is any realistic systems in which these processes are manifest and can be tuned~\cite{Damle2015}. The dipolar kagome ice was first realized in an artificial kagome array of nanomagnets with the nanosized magnetic bars playing the role of Ising spins~\cite{Rougemaille2011}. Direct imagining reveals evidence for the magnetic charge ordered phase and the $\sqrt{3}\times\sqrt{3}$ phase in this system~\cite{Canals2016}. Meanwhile, it possesses out-of-equilibrium features unique to nanomagnetic systems ~\cite{Chioar2014kinetic}. 

On the material front, rare-earth based tripod kagome magnets Mg$_2$R$_3$Sb$_3$O$_{14}$ (R$=$Dy, Ho) are thought to be the material incarnations of the dipolar kagome spin ice~\cite{Dun2016,Dun2017}. The rare earth ions form ABC stacked kagome planes in this family of materials. For compounds made from Kramers magnetic ions such as Dy$^{3+}$, the elementary degrees of freedom are their lowest energy crystal field doublet, which map onto Ising spins. The large magnetic moment carried by the Ising spins, along with the relatively weak super exchange interactions, imply that the spins interact predominantly through the magnetic dipole-dipole interaction. The material thus can be modeled as a dipolar kagome spin ice as a first approximation. Thermodynamic measurements and neutron scattering have provided experimental evidence for the magnetic charge order in Mg$_2$Dy$_3$Sb$_3$O$_{14}$~\cite{Paddison2016}, which is consistent with the dipolar kagome spin ice picture. 

Quantum effects set in when the rare earth ions are non-Kramers ions such as Ho$^{3+}$~\cite{Dun2019}. As the time reversal symmetry no longer protects the degeneracy of the crystal field doublet, the doublet in an isolated ion would split into two quasi-degenerate, non-magnetic singlets due to the low crystal field symmetry. Such crystal field effect can be viewed as an effective transverse field acting on the aforementioned Ising spins~\cite{Wang1968}. The quantum fluctuations brought in by the crystal field effect competes with the magnetic dipole-dipole interaction, thereby offering another handle to tune the physics of dipolar kagome spin ice, and, in particular, the melting processes of the $\sqrt{3}\times\sqrt{3}$ magnetic ground state.

These considerations motivate us to explore the following minimal model that captures the competition between the long range magnetic dipole-dipole interaction and the quantum fluctuations,
\begin{align}
H &= D\sum_{i>j}\frac{\hat{z}_i\cdot\hat{z}_j-3(\hat{z}_i\cdot\hat{r}_{ij})(\hat{z}_j\cdot\hat{r}_{ij})}{(r_{ij}/r_\mathrm{nn})^3}\sigma^z_i\sigma^z_j 
\nonumber\\
&- h\sum_{i}\sigma^x_i,
\label{eq:hamiltonian}
\end{align}
where $\sigma^{x,y,z}_{i}$ are Pauli matrices that describe the effective Ising spin at site $i$. $r_{ij}$ is the spatial distance between the kagome site $i$ and $j$, and $\hat{r}_{ij}$ is the unit vector that points from $i$ to $j$. $r_\mathrm{nn}$ is the distance between nearest neighbors. $D>0$ sets the energy scale of the dipole interaction, while $h$ is the strength of the transverse field. We assume $h>0$ without loss of generality. 

Unit vector $\hat{z}_i$ describes the direction of the local $\sigma^z_i$ axis. For simplicity, we set $\hat{z}_i$ to be in the kagome plane and pointing toward the center of the up triangles (Fig.~\ref{fig:cartoon}a) similar to the previous studies~\cite{Moeller2009,Chern2011,Chern2012}. This choice is natural from a purely geometric point of view and simplifies the physics of the model~\footnote{The other natural choice is to make $\hat{z}_i$ in the kagome plane and pointing away from the center of the up triangles. These two choices are related by flipping the sign of $\sigma^z_i$ in the Hamiltonian and therefore physically equivalent.}. However, in tripod kagome materials, $\hat{z}_i$ cants away from the kagome plane by about $26^\circ$~\cite{Dun2016,Dun2017,Paddison2016}, which has significant impact on the magnetic dipole-dipole interactions. We shall return to this point in Sec.~\ref{sec:discussion}

Eq.~\ref{eq:hamiltonian} possesses all the lattice symmetries. Crucially, it also possesses the time reversal symmetry despite the transverse field term. This is due to the fact that the Pauli matrices $\sigma^{x,y,z}_i$ describe the two lowest crystal field levels of a rare earth ion rather than a physical $S=1/2$ spin. While $\sigma^z_{i}$ changes sign under the time reversal, $\sigma^{x}_i$ remains invariant.~\footnote{Taking Ho$^{3+}$ as an example, the two lowest crystal field levels are approximately linear superpositions of the states \unexpanded{$|J=8,m_J=\pm8\rangle$}. We define the Pauli matrices as follows: $\sigma^z$ takes eigenvalue $\pm1$ in the states \unexpanded{$|J=8,m_J=\pm8\rangle$}, while $\sigma^x$ exchange these two states. One can easily check that the so defined $\sigma^x$ matrix is invariant under time reversal.}

In this work, we investigate the thermodynamic phase diagram of Eq.~\eqref{eq:hamiltonian} by performing a quantum Monte Carlo (QMC) simulation\textcolor{blue}{.} The results are summarized in the phase diagram (Fig.~\ref{fig:phase_diag}). At low temperature, the system hosts the $\sqrt{3}\times\sqrt{3}$ phase over the entire parameter window of simulation, $h/D \in [0,0.65]$. As the temperature increases, the $\sqrt{3}\times\sqrt{3}$ phase melts through the two aforementioned pathways. For weak transverse field $h/D<0.5$, the $\sqrt{3}\times\sqrt{3}$ order melts through the intermediate magnetic charge ordered phase, which is connected to the classical limit~\cite{Moeller2009,Chern2011,Chern2012}. For stronger transverse field $0.5<h/D<0.65$, the $\sqrt{3}\times\sqrt{3}$ phase melts through the floating KT phase, similar to the kagome Ising model with first and second neighbor interactions~\cite{Wolf1988,Takagi1993,Wills2002}. Our results thus reveal the interesting prospect of tuning the multistep melting of complex magnetic orders through quantum fluctuations.

The rest of this work is organized as follows. In Sec.~\ref{sec:landau}, we perform a Landau theory analysis of the melting processes of the $\sqrt{3}\times\sqrt{3}$ phase, which provides a natural framework for organizing the QMC data. In Sec.~\ref{sec:method}, we explain the QMC algorithms and the Monte Carlo observables employed in this work. In Sec.~\ref{sec:results}, we present a detailed analysis of the QMC data, which forms the basis of the phase diagram Fig.~\ref{fig:phase_diag}. Finally, in Sec.~\ref{sec:discussion}, we point out a few open problems that are worth exploring in future.

\section{Landau theory \label{sec:landau}}

Before embarking on the QMC simulation of Eq.~\eqref{eq:hamiltonian}, we set the stage by performing a Landau theory analysis~\cite{Domany1978,Domany1979,Damle2015}. As we shall see, the Landau theory provides a natural framework to understand and organize the QMC results.

To this end, we construct the relevant order parameters. The order parameter for the magnetic charge order is defined as the magnetic charge imbalance between the up and down triangles,
\begin{subequations}
\begin{align}
m = \frac{1}{2N}(\sum_{\alpha\in\bigtriangleup}Q_\alpha - \sum_{\alpha\in\bigtriangledown}Q_\alpha).
\end{align}
where $\alpha$ labels the triangles of the kagome lattice. $Q_\alpha$ is the magnetic charge enclosed by the triangle $\alpha$. The first and second summation runs over all the up and down triangles, respectively. $N$ is the number of lattice sites. We may express $Q_\alpha$ in terms of $\sigma^z_i$: 
\begin{align}
m &= \frac{1}{2N}(\sum_{\alpha\in\bigtriangleup}\sum_{i\in\alpha}\hat{z}_i\cdot\mathbf{m}_i-\sum_{\alpha\in\bigtriangledown}\sum_{i\in\alpha}(-\hat{z}_i\cdot\mathbf{m}_i))
\nonumber\\
& = \frac{1}{2N}(\sum_{\alpha\in\bigtriangleup}\sum_{i\in\alpha}\sigma^z_i+\sum_{\alpha\in\bigtriangledown}\sum_{i\in\alpha}\sigma^z_i)
= \frac{1}{N}\sum_{i}\sigma^z_i.
\label{eq:m_def}
\end{align}
\end{subequations}
In the first line, we have inserted the definition of magnetic charge, i.e. the total magnetization flux flowing \emph{into} the enclosed area. $\mathbf{m}_i$ is the magnetic moment at site $i$. In the second line, we have used $\mathbf{m}_i = \sigma^z_i \hat{z}_i$. We see that the order parameter $m$ is essentially the uniform magnetization measured in the \emph{local} spin $\sigma^z_i$ axis. This can be understood as follows. In the magnetic charge ordered phase, the static magnetic moments form a staggered, all-in-all-out pattern (Fig.~\ref{fig:cartoon}c). As the local $\sigma^z_i$ axis form the same staggered, all-in-all-out pattern (Fig.~\ref{fig:cartoon}a), projecting the magnetic moment to the local $\sigma^z_i$ axis yields a uniform value of $\sigma^z_i$. From Eq.~\eqref{eq:m_def}, it is easy to see $m\to -m$ under time reversal and transforms trivially under lattice translations. Both properties agree with the symmetry properties of the magnetic charge order.

The $\sqrt{3}\times\sqrt{3}$ phase breaks both time reversal symmetry and the lattice translation symmetry. It inherits the same magnetic charge order from the magnetic charge ordered phase, which is captured by the order parameter $m$ (see Fig.~\ref{fig:cartoon}\textcolor{blue}{d} for the magnetic charge distribution in the $\sqrt{3}\times\sqrt{3}$ phase). The spontaneous breaking of the translation symmetry is captured by another order parameter,
\begin{align}
\psi = \frac{1}{N}\sum_{i} \sigma^z_i e^{i\mathbf{Q}\cdot\mathbf{r}_i}.
\end{align}
$\mathbf{Q} = 2\mathbf{K}$ is the characteristic wave vector associated with the $\sqrt{3}\times\sqrt{3}$ magnetic unit cell, where $\mathbf{K} = (2\pi/(3r_\mathrm{nn}),0)$ is the lattice wave vector corresponding to the K point of the first Brillouin zone. $r_\mathrm{nn}$ is the distance between nearest neighbor kagome sites. $\mathbf{r}_i$ is the position vector of the kagome site $i$. Note the value of $\mathbf{r}_i$ depends on the origin of the coordinate system. Our choice is shown in  Fig.~\ref{fig:cartoon}a. $\psi\to -\psi$ under the time reversal, $\psi\to\psi^\ast$ under the site inversion, and $\psi \to \exp(i\mathbf{Q}\cdot\mathbf{R})\psi$ under the translation by lattice vector $\mathbf{R}$. These transformation properties are consistent with the symmetry properties of the $\sqrt{3}\times\sqrt{3}$ phase.

Provided that no other order parameters are present, the Landau free energy $F$ is a polynomial of  $m$ and $\psi$. $F$ contains terms such as $m^p|\psi|^q$, $m^p|\psi|^q\cos(q\theta)$, and $m^p|\psi|^q\sin(q\theta)$, where $p,q$ are integers and $\theta$ is the complex phase angle of $\psi$. First of all, the lattice inversion symmetry forbids $m^p|\psi|^q\sin(q\theta)$. Secondly, the translation symmetry requires $q$ to be multiples of 3 in $m^p|\psi|^q\cos(q\theta)$.  Thirdly, the time reversal symmetry requires $p+q$ to be an even number. Combining all of these symmetry requirements yields,
\begin{subequations}
\label{eq:landau}
\begin{align}
F = F_m + F_\psi + F_{m,\psi}.
\end{align}
$F_m$ is the Landau free energy for $m$:
\begin{align}
F_m = \alpha_m m^2 + \beta_m m^4.
\end{align}
$F_\psi$ is the Landau free energy for $\psi$:
\begin{align}
F_\psi = \alpha_\psi |\psi|^2 + \beta_\psi |\psi|^4 + \gamma_\psi |\psi|^6 -\delta_\psi |\psi|^6\cos(6\theta),
\end{align} 
$F_{m,\psi}$ describes the coupling between the two order parameters~\cite{Damle2015}:
\begin{align}
F_{m,\psi} = -g m |\psi|^3\cos(3\theta) + g' m^2|\psi|^2.
\end{align}
\end{subequations}
We have omitted in Eq.~\eqref{eq:landau} higher order terms that are inessential to the present discussion.

We are interested in the phase transitions driven by $\alpha_{m},\alpha_{\psi}$. To this end, we need to fix the sign of all the other coefficients. We set $\beta_m,\beta_\psi,\gamma_\psi>0$. We also assume $|\delta_\psi|$, $|g|$, $|g'|$ are sufficiently small to ensure $F$ is bounded from below. Since the sign of $g$ can be absorbed into the order parameter $m$, we set $g>0$ without loss of generality. $g'$ describes the mutual enhancement ($g'<0$) or suppression ($g'>0$) of the order parameters. Since it doesn't change the qualitative features of the mean field phase diagram, we shall omit it for now and return to it later. To fix the sign of $\delta_\psi$, we observe that, with our coordinate system (Fig.~\ref{fig:cartoon}a), the complex phase of the order parameter $\theta = n\pi/3$, where $n = 0, 1, 2,\cdots 5$, in the $\sqrt{3}\times\sqrt{3}$ magnetic ground state (Fig.~\ref{fig:cartoon}d). This suggests $\delta_\psi>0$. By contrast, $\delta_\psi<0$ would favor $\theta  = n\pi/3+\pi/6$, which correspond to the partially disordered states~\cite{Takagi1993,Wills2002} that are not observed in this work (Fig.~\ref{fig:cartoon}e).

\begin{figure}
\includegraphics[width = \columnwidth]{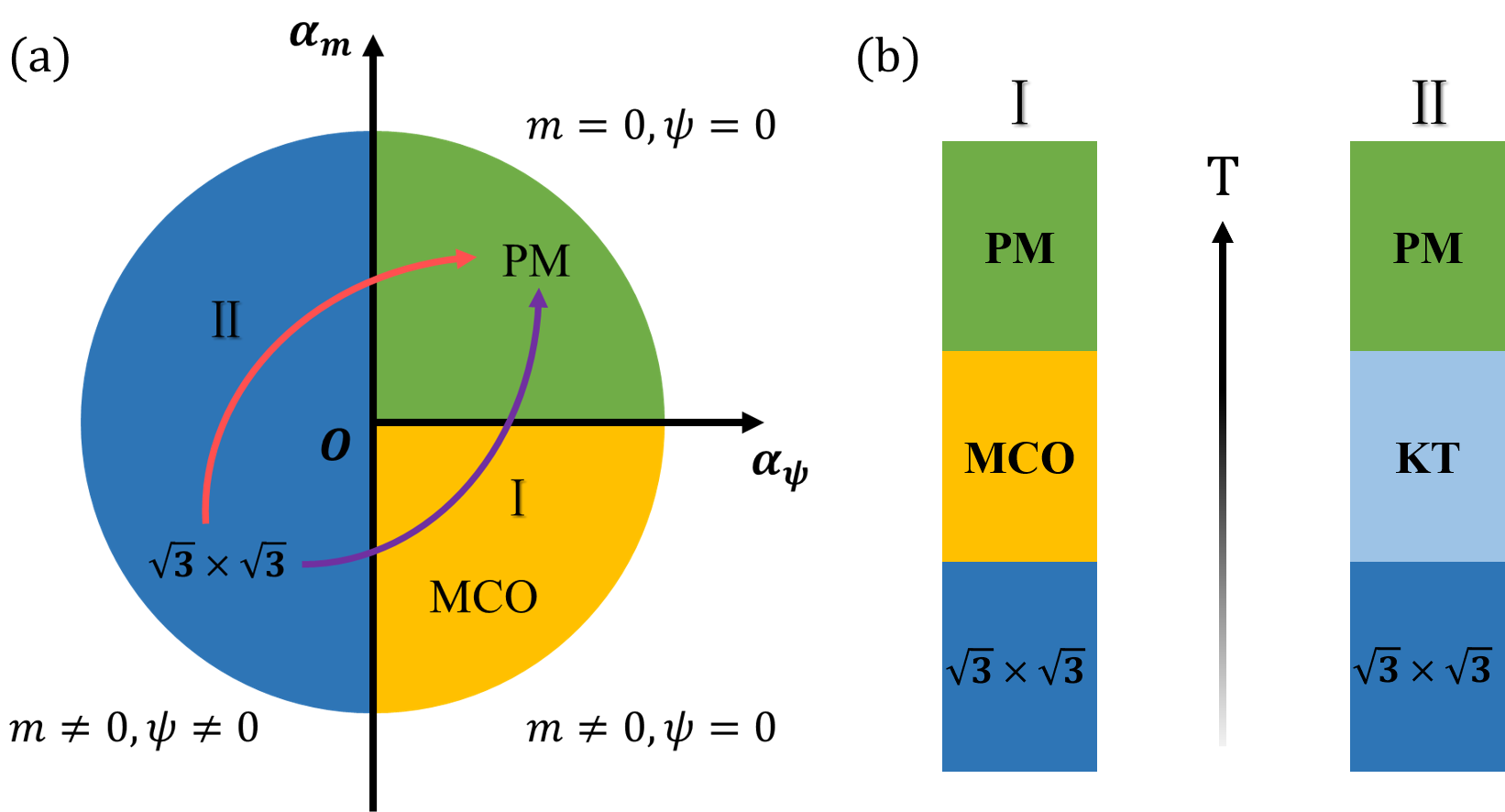}
\caption{(a) Landau mean field theory phase diagram as a function of the parameters $\alpha_{m,\psi}$. Arrows show the two generic pathways (I and II) that the $\sqrt{3}\times\sqrt{3}$ phase can melt. (b) The sequence of thermal phases corresponding to the pathway I and II. Note the pathway II shows a floating Kosterlitz-Thouless phase that is beyond the mean field theory but expected to exist. 
\label{fig:landau_theory}}
\end{figure}

Fig.~\ref{fig:landau_theory}a presents the mean field phase diagram as a function of $\alpha_m,\alpha_\psi$, which contains three phases: the paramagnetic phase ($m = 0, \psi = 0$), the magnetic charge ordered phase ($m \neq 0, \psi = 0$), and the $\sqrt{3}\times\sqrt{3}$ phase ($m \neq 0, \psi \neq 0$). Note phase with $m=0$ but $\psi \neq 0$ does not appear in that $\psi\neq0$ breaks time reversal symmetry, which necessarily induces a finite $m$.

The mean field theory predicts two \emph{generic} melting pathways that connect the paramagnetic phase and the $\sqrt{3}\times\sqrt{3}$ phase: either through an intermediate magnetic charge ordered phase, or through a direct, continuous phase transition. While the former pathway is consistent with the behavior of the classical kagome spin ice~\cite{Moeller2009,Chern2011,Chern2012}, the latter pathway cannot occur generically. To see this, we note that the latter pathway is driven by $F_\psi$ (Eq.~\eqref{eq:landau}), which resembles the Landau free energy of the six-state clock model. The two dimensional six-state clock model exhibits the floating KT phase between the fully ordered phase and the paramagnetic phase~\cite{Jose1977,Challa1986}. Therefore, the pathway II must feature the floating KT phase, where $\psi$ shows algebraic long range correlation. Fig.~\ref{fig:landau_theory}b presents the ``corrected" sequence of phases for pathways I \& II.

We may also deduce the aforementioned melting pathways by an analogy to a \emph{generalized} six-state clock model that contains three independent, symmetry allowed interactions~\cite{Cardy1980,Dorey1999,Chern2012}. The $\sqrt{3}\times\sqrt{3}$ phase has 6 symmetry-related domains, which can be thought of as the 6 states of a clock spin. The symmetry of the Landau theory Eq.~\eqref{eq:landau} is the symmetry group of a hexagon, $D_6 = S_3\times Z_2$, where $S_3$ permutes the three domains that share the same value of order parameter $m$ and $Z_2$ is the time reversal symmetry. This symmetry group coincides with the symmetry group of the generalized six-state clock model. On one hand, the $D_6$ symmetry of the six-state clock model may spontaneously break through the floating KT phase. On the other hand, one may expect the $D_6$ group can be first broken down to its subgroup $S_3$ (magnetic charge ordered phase) and then become fully broken ($\sqrt{3}\times\sqrt{3}$ phase), namely the symmetry sequence $D_6\to S_3\to I$. 

Note the generalized six-state clock model also admits the symmetry sequence $D_6\to Z_2\to I$~\cite{Cardy1980,Dorey1999,Chern2012}. In the present context, this sequence would require an intermediate phase that preserves the time reversal symmetry but breaks the lattice symmetry. Such an intermediate phase does not occur in the Landau theory analysis Eq.~\eqref{eq:landau} and is not observed in the QMC simulations of Eq.~\eqref{eq:hamiltonian}. It is also important to bear in mind that the effective theory Eq.~\ref{eq:landau} is not exactly mapped to a generalized six-state clock model~\cite{Damle2015}. We shall comment more on this point in Sec.~\ref{sec:interm_h}.

To summarize, we expect Eq.~\eqref{eq:hamiltonian} to exhibit two distinct sequences of thermal phase transitions as the $\sqrt{3}\times\sqrt{3}$ phase melts: either through an intermediate magnetic charge ordered phase, or through a floating KT phase. The QMC-constructed phase diagram of Eq.~\eqref{eq:hamiltonian} (Fig.~\ref{fig:phase_diag}) shows that the former sequence occurs near the classical limit, while the latter sequence occurs for larger quantum fluctuations. We shall present a detailed analysis of our QMC data in Sec.~\ref{sec:results}.

\section{Method \label{sec:method}}

\subsection{Algorithm}

We perform QMC simulations of Eq.~\ref{eq:hamiltonian} based on the standard second order Trotter decomposition. We set the discretization time $h\delta\tau = 0.02$, with the number of imaginary time slices $N_{\tau} = \beta / \delta \tau$. Our choice of $h\delta\tau$ is sufficiently small so that the systematic discretization error on all observables is smaller than the statistical one. Comparing to QMC schemes free of discretization errors, such as stochastic series expansion~\cite{Sandvik2003}, the present scheme allows for a straightforward extension of classical non-local updates to the quantum realm~\cite{Henry2014}, which are essential for an effective sampling at small transverse field. We shall return to this point momentarily.

We use a cluster of $L\times L$ primitive unit cells subject to the periodic boundary condition and treat the long range magnetic dipole-dipole interaction with the Ewald summation~\cite{Grzybowski2000}. We set $L$ to be multiples of 3 so that the cluster accommodates the $\sqrt{3}\times\sqrt{3}$ magnetic unit cell. The long range magnetic dipole-dipole interaction introduces significant geometric frustration to the model. Furthermore, it renders the QMC simulation computationally more expensive comparing to similar models with short range interactions. Both factors limit the accessible system size $L$ to 24 or smaller,  and the accessible temperature $k_BT/D$ to $10^{-2}$.

The Trotter decomposition maps Eq.~\eqref{eq:hamiltonian} to an effective three dimensional classical Ising model. On one hand, the interactions along the imaginary time direction are nearest neighbor and ferromagnetic, which implies that we may use conventional cluster update along the world line direction. On the other hand, the interactions in the spatial directions are long range and frustrated. In particular, in the classical limit ($h/D=0$), the loop updates are necessary for an effective sampling~\cite{Barkema1998,Moeller2009,Chern2011,Chern2012}. These observations motivate us to adopt two complementary update schemes, namely the ``line" update~\cite{Nakamura2008} and the ``membrane" update~\cite{Henry2014}. 

In a line update~\cite{Nakamura2008}, we choose a spin at random and then perform the Swendsen-Wang or Wolff cluster update along the world line of the chosen spin. The line update eliminates the dynamical freezing due to strong ferromagnetic couplings in the imaginary time direction, which would otherwise render the single spin flip update inefficient. 

The membrane update~\cite{Henry2014} may be viewed as an extension of the loop update~\cite{Barkema1998} to quantum models. The membrane update proceeds as follows. We first choose a time slice at random and construct a closed loop of spins with staggered values of $\sigma^z_i$ in the said time slice. To this end, we use both long loops, in which the loop head closes on the starting point, and short loops, in which the loop head hits on an already constructed loop segment, whereupon the dangling tail is discarded. In the next step, we grow the loop to the adjacent time slices akin to a Wolff cluster. This forms the spin membrane. Finally, we flip the spins in the said membrane according to the Metropolis rule. As the membrane update is not irreducible by itself, it must be complemented with the line update. 

In this work, we employ both update schemes for weak transverse field $h/D<0.5$, and only the line update for stronger transverse field $h/D\ge0.5$ in that the membrane update becomes less effective as $h/D$ increases. For $h/D<0.5$, each Monte Carlo step (MCS) consists of 5 lattice sweeps of line updates followed by 2-5 membrane updates. In each membrane update, we carry out $O(L)$ attempts to build a long loop and $O(L^2)$ attempts to build a short loop. For $h/D>0.5$, each MCS consists of 1 lattice sweep of line updates. We parallelize the Markov chain using at least $8 \times 10^{3}$ thermalization MCS followed by $O(10^{4})$ measurement MCS for each independent Markov chain, which resulted in a total of $O(10^6)$ thermalization and $O(10^6)$ measurement steps per parameter set. 

\subsection{Observables}

We employ the following Monte Carlo observables to detect the phases and phase transitions.  We estimate the specific heat per site $C_{v}$ using the approximant~\cite{Suzuki1976}:
\begin{align}
 C_{v} \approx \frac{1}{N} \beta^2 \left. \frac{\partial^2}{\partial \beta^2} \ln Z_\mathrm{Trotter}(\frac{\beta}{N_\tau}) \right|_{N_{\tau} = \text{const.}},
 \label{eq:cv_approx}
\end{align}
where $Z_\mathrm{Trotter}(\beta/N_\tau)$ is the approximate partition function of Eq.~\eqref{eq:hamiltonian} for the Totter discretization time $\delta\tau = \beta/N_\tau$ and the number of time slices $N_\tau$. This specific heat approximant is known to show a spurious peak at the temperature scale $k_BT_{\text{spurious}} \sim h/N_{\tau}$ due to the Trotter discretization error~\cite{Fye1987}. This spurious peak makes the specific heat estimate unreliable at very low temperature.

We characterize the magnitude of the order parameters $m$ and $\psi$ through,
\begin{align}
\langle m^2\rangle &=\overline{O}_m \overset{\mathrm{def}}{=} \overline{\frac{1}{N_\tau}\sum_\tau\left(\frac{1}{N}\sum_{i}s_{i,\tau}\right)^2}, \nonumber\\
\langle |\psi|^2\rangle &= \overline{O}_\psi \overset{\mathrm{def}}{=} \overline{\frac{1}{N_\tau}\sum_\tau\left|\frac{1}{N}\sum_{i}s_{i,\tau}e^{i\mathbf{Q}\cdot\mathbf{r}_i}\right|^2}.
\end{align}
where $s_{i,\tau}=\pm1$ is the Ising variable of the effective three-dimensional Ising model in the QMC simulation. $i,\tau$ label the site and the time slices, respectively. $\overline{O}$ stands for the Monte Carlo average of the observable $O$.

We define the Binder ratios as follows~\cite{Binder1981aa,Sandvik2010ComputationalStudies},
\begin{align}
\frac{\langle m^4\rangle}{\langle m^2\rangle^2} \overset{\mathrm{def}}{=} \frac{\overline{O^2_m}}{(\overline{O}_m)^2},\quad
\frac{\langle |\psi|^4\rangle}{\langle |\psi|^2\rangle^2} \overset{\mathrm{def}}{=}  \frac{\overline{O^2_\psi}}{(\overline{O}_\psi)^2}.
\end{align}
The above definition uses the moments of distribution of the Monte Carlo observables $O_m$ and $O_\psi$ in the effective \emph{classical} system. Alternatively, one could construct Binder ratio using the \emph{quantum} average of the high order moments of the order parameters. These definitions have similar asymptotic system size dependence. In particular, the crossing point analysis would asymptotically yield the same critical temperature although the value at the critical point may differ. 

$\theta$, the complex phase angle of $\psi$, contains important information about the magnetic order. Specifically,  $\theta = n\pi/3$, $n = 0,1,2,\cdots 5$, in the $\sqrt{3}\times\sqrt{3}$ magnetic ground state (Fig.~\ref{fig:cartoon}d), whereas $\theta = n\pi/3 + \pi/6$ in the partially disordered state (Fig.~\ref{fig:cartoon}e). We distinguish these two possibilities by using the anisotropy measure:
\begin{align}
\frac{\langle |\psi|^6\cos(6\theta)\rangle}{\langle |\psi|^6\rangle} = \frac{\displaystyle\overline{\mathrm{Re}\left(\frac{1}{N}\sum_{i}s_{i,\tau=0}e^{i\mathbf{Q}\cdot\mathbf{r}_i}\right)^6}}{\displaystyle\overline{\left|\frac{1}{N}\sum_{i}s_{i,\tau=0}e^{i\mathbf{Q}\cdot\mathbf{r}_i}\right|^6}}.
\end{align}
In particular, the anisotropy measure  approaches 1 and $-1$ in the $\sqrt{3}\times\sqrt{3}$ state and partially disordered state, respectively.

Finally, we have also measured the spin structure factor to detect any potential magnetic ordering with magnetic unit cells other than $\sqrt{3}\times\sqrt{3}$. We have only found signatures of magnetic ordering with characteristic wave vector  $\mathbf{Q} = 2\mathbf{K}$, rendering such possibilities unlikely. 

\section{Results \label{sec:results}}

In this section, we present a detailed analysis of our QMC data. In Sec.~\ref{sec:weak_h}, we focus on the thermal phase transitions at $h/D=0.25$, which is representative for the weak transverse field regime ($h/D<0.5$). Next, in Sec.~\ref{sec:strong_h}, we turn to the strong transverse field regime ($h/D>0.5$) and focus on the representative case with $h/D=0.6$. Finally, in Sec.~\ref{sec:interm_h}, we discuss the intermediate case $h/D=0.5$, which separates the distinct behaviors at weak and strong transverse fields.

\subsection{Weak transverse field $h/D<0.5$\label{sec:weak_h}}

\begin{figure*}
\includegraphics[width = \textwidth]{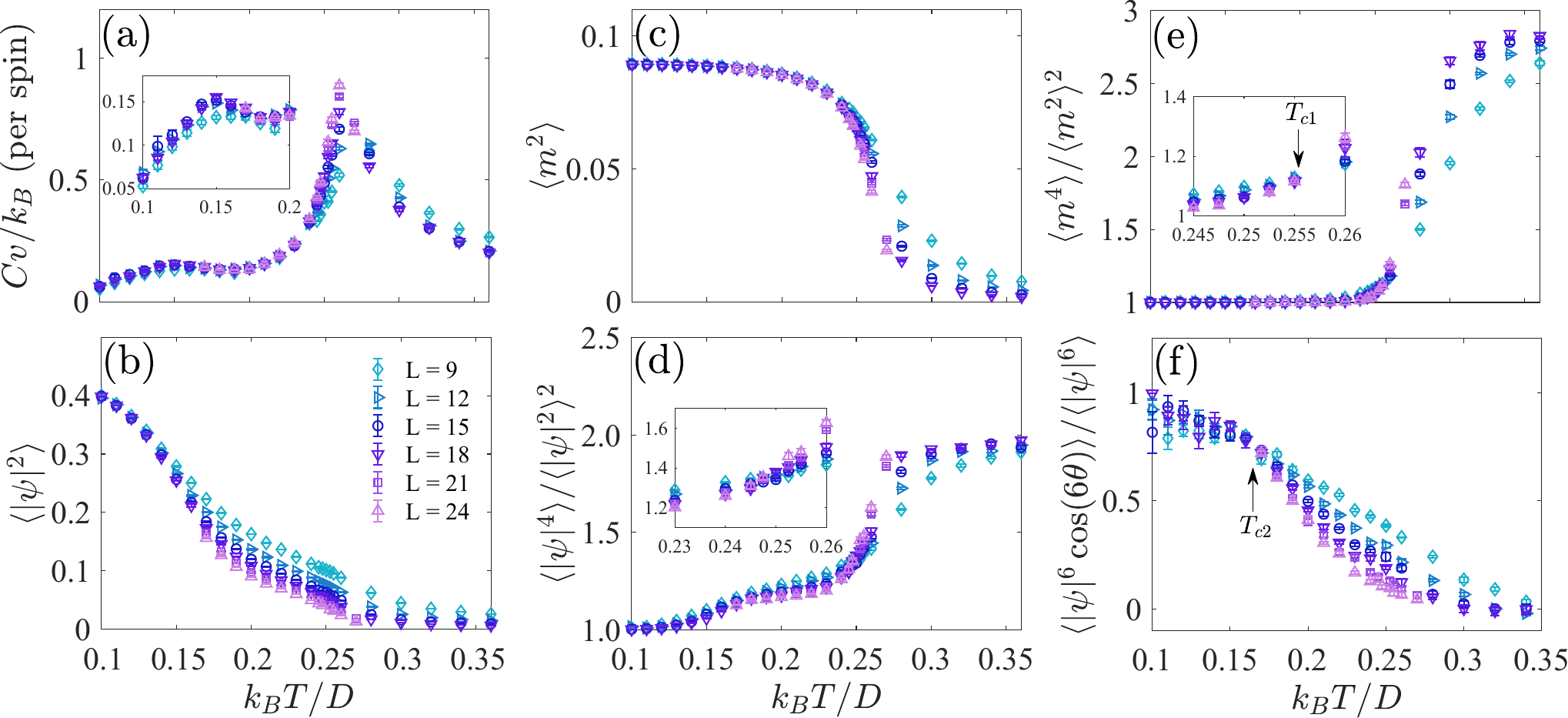}
\caption{Specific heat $C_{v}$ (a), magnitude of the order parameter for the $\sqrt{3}\times\sqrt{3}$ order $\langle |\psi|^2\rangle$ (b), magnitude of the order parameter for the magnetic charge order $\langle m^2\rangle$ (c), their respective Binder ratios (d,e), and 6-fold anisotropy measure of $\psi$ (f) as functions of temperature $T$ at $h/D = 0.25$. Data for different system sizes $L$ are in different colors and symbols. Insets in (a), (d), and (e) present enlarged views of the respective temperature windows. Arrows in (e) and (f) mark the estimated locations of the high temperature ($T_{c1}$) and the low temperature phase transition ($T_{c2}$).  
\label{fig:h_0p25}}
\end{figure*}

\begin{figure*}
\includegraphics[width = 0.75\textwidth]{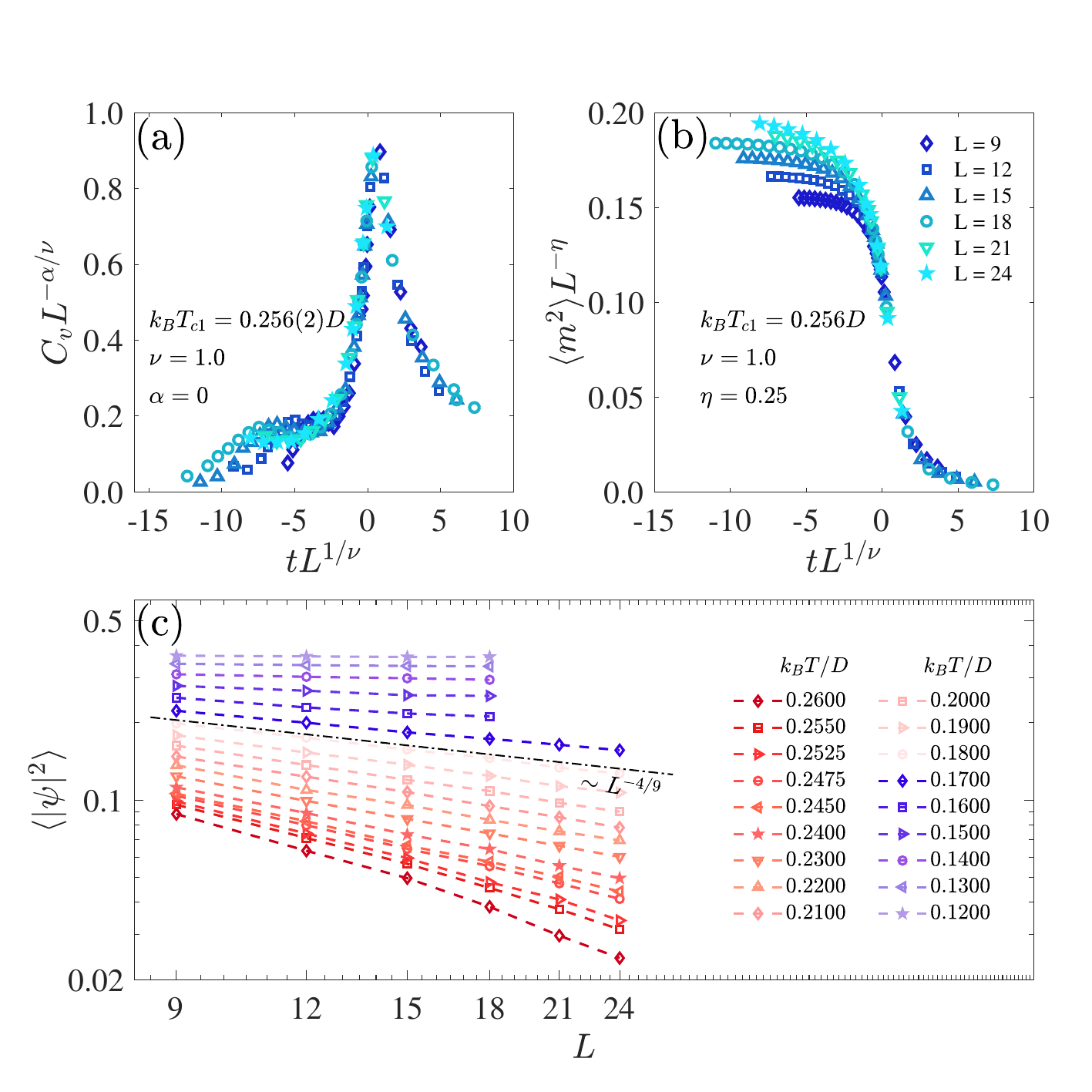}
\caption{(a) Data collapse for the specific heat for the high temperature transition at $T_{c1}$ using the critical exponents from the Ising universality class. $t = (T-T_{c1})/T_{c1}$. (b) Data collapse for $\langle m^2\rangle$ using the exponents from the Ising universality class. (c) Log-log plot of $\langle |\psi|^2\rangle$ as a function of system size $L$ at different temperatures. Data fall into the magnetic charge ordered phase and the $\sqrt{3}\times\sqrt{3}$ phase are shaded in red and purple, respectively. Dashed line shows the expected $L^{-4/9}$ scaling at $T_{c2}$ deduced from the effective dimer model assuming no charge defects are present in the system.
\label{fig:h_0p25_data_collapse}}
\end{figure*}

Fig.~\ref{fig:h_0p25} shows a temperature scan of the phase diagram at constant transverse field $h/D = 0.25$, which is representative for the weak transverse field regime. The specific heat $C_v$ shows a sharp peak at $k_BT/D \approx 0.26$ (Fig.~\ref{fig:h_0p25}a). Meanwhile, $\langle m^2\rangle$ increases rapidly at about the same temperature scale, and the onset of $\langle m^2\rangle$ is more abrupt for larger system size $L$ (Fig.~\ref{fig:h_0p25}c). Both behaviors point to a second order phase transition from the paramagnetic phase to the magnetic charge ordered phase. We estimate the transition temperature $T_{c1}$ by plotting the Binder ratio of $m$ for various system sizes. Inspecting the crossing point of the Binder ratio for $L=21$ and $L=24$ yields the estimate $k_BT_{c1}/D \approx 0.256(2)$ (Fig.~\ref{fig:h_0p25}e) where the number in the brackets indicates the uncertainty in the last digit.

Given the symmetry of the order parameter $m$, the magnetic charge ordering transition is expected to be in the Ising universality class~\cite{Moeller2009,Chern2011}. We are able to collapse the data for $C_v$ and $\langle m^2\rangle$ by using exponents from the two-dimensional Ising transition (Fig.~\ref{fig:h_0p25_data_collapse}a\&b). In particular, the $T_{c1}$ determined from the data collapse agrees reasonably well with the one estimated based on the Binder ratio.

Having established the transition from the paramagnetic phase to the magnetic charge ordered phase, we turn to the order parameter $\psi$. $\langle |\psi|^2\rangle$ shows rich behavior as a function of temperature (Fig.~\ref{fig:h_0p25}b). It starts developing at about the same temperature scale as $T_{c1}$ and shows a kink at a lower temperature scale $k_BT/D \approx 0.16$. Accompanying this kink, the specific heat $C_v$ starts developing a bump below this temperature. The height of the bump does not increase significantly as $L$ increases (Fig.~\ref{fig:h_0p25}a, inset). Meanwhile, the Binder ratio of $\psi$ for different $L$ cross near $T_{c1}$, and converge again at a lower temperature near $k_BT/D \approx 0.15$ (Fig.~\ref{fig:h_0p25}e).

We interpret these results as follows. The onset of $\langle|\psi|^2\rangle$ near $T_{c1}$ reflects the \emph{pseudo}-critical fluctuations of $\psi$ induced by the magnetic charge order. Moreover, the converging of the Binder ratio, along with the specific heat bump at the lower temperature scale, reflect a \emph{pseudo}-KT transition from the magnetic charge ordered phase to $\sqrt{3}\times\sqrt{3}$ phase at $T_{c2}<T_{c1}$. The physics behind pseudo-critical fluctuations of $\psi$ and the pseudo-KT transition is already understood in the context of classical limit; here, we briefly reproduce the argument for the sake of completeness~\cite{Moeller2009,Chern2011}.

Consider a perfect magnetic charge order in which the up (down) triangles carry $+1$ ($-1$) magnetic charge (Fig.~\ref{fig:cartoon} b\&{}c). This effectively establishes an ice rule on the spin configurations similar to the classical spin ice~\cite{Gingras2011} --- an up triangle must have 2 spins pointing inward, and 1 spin outward; likewise, a down triangle must have 1 spin pointing outward, and 2 spins inward. The system fluctuates in this restricted manifold of spin configurations, giving rise to an algebraic spin correlation~\cite{Moeller2009,Chern2011}. Furthermore, the transition from the magnetic charge ordered phase to the $\sqrt{3}\times\sqrt{3}$ phase is of the KT universality if the ice rules are strictly enforced~\cite{Alet2005,Alet2006,Moeller2009,Chern2011}.

In reality, defect triangles that violate the ice rules always appear with finite density. Therefore, the aforementioned algebraic spin correlation is cut off by a crossover length scale set by the average distance between defect triangles. The spins are short range correlated above the crossover length scale. In addition, the $\sqrt{3}\times\sqrt{3}$ ordering transition crosses over from the KT universality to the three-state Potts universality above the said length scale~\cite{Chern2011}. 

Now, given the limited system size $L$, we expect that $L$ is smaller than the crossover length scale. The algebraic-like spin correlation produces the pseudo-critical fluctuations in $\psi$, which in turn are responsible for the enhancement of $\langle|\psi|^2\rangle$ below $T_{c1}$ in finite-size systems. Phenomenologically, the enhancement in the fluctuations of $\psi$ due to $m$ is captured by the coupling $g' m^2|\psi|^2$ in the Landau free energy Eq.~\eqref{eq:landau} with $g'<0$. The crossing of Binder ratio $\langle |\psi|^4\rangle / \langle |\psi|^2\rangle^2$ at $T_{c1}$ is also attributed to this coupling. 

The small system size implies the transition from the magnetic charge ordered phase to the $\sqrt{3}\times\sqrt{3}$ phase is controlled by the KT universality. The severe finite size effect makes it difficult to determine its critical temperature $T_{c2}$ even for the classical model~\cite{Moeller2009,Chern2011,Chern2012}. In this work, we use the anisotropy measure of $\psi$~\cite{Isakov2003} (Fig.~\ref{fig:h_0p25}f). It is positive throughout and approaches $1$ as temperature decreases. This shows that the system enters the $\sqrt{3}\times\sqrt{3}$ phase as opposed to the partially disordered state. The anisotropy measure data for different $L$ cross at approximately $k_BT_{c2}/D \approx 0.17(1)$, which we take to be the critical temperature.

We further test the picture of pseudo-KT transition by examining the correlation of $\psi$ in the magnetic charge ordered phase. We expect $\psi$ exhibit algebraic like correlation in a large temperature window of the magnetic charge ordered phase. This would imply the finite-size scaling $\langle |\psi|^2\rangle\sim L^{-\eta}$, where $\eta$ is the anomalous dimension of $\psi$. The log-log plot of $\langle |\psi|^2 \rangle$ as a function of $L$ seems to be consistent with this idea (Fig.~\ref{fig:h_0p25_data_collapse}c): in a broad temperature window between $T_{c1}$ and $T_{c2}$, the log-log plot resembles a straight line in the limited range of $L$. In particular, had there been no defect triangles, we expect $\langle|\psi|^2\rangle\sim L^{-4/9}$ at $T_{c2}$~\cite{Alet2005,Alet2006}. The observed slope near $T_{c2}$ is fairly close to this scaling. 

To recapitulate, at $h/D = 0.25$, the system enters first the magnetic charge ordered phase and then the $\sqrt{3}\times\sqrt{3}$ phase as temperature decreases, i.e. the pathway I shown in Fig.~\ref{fig:landau_theory}b. The magnetic charge ordering transition is of Ising universality. The $\sqrt{3}\times\sqrt{3}$ magnetic ordering transition is expected to be of three-state Potts universality in the thermodynamic limit. Here, it exhibits pseudo-KT universality due to the finite size effect.

The thermal transitions at other values of $h/D<0.5$ shows similar behaviors. Employing the Binder ratio of $m$ and the anisotropy measure for $\psi$, we estimate the phase boundaries between the paramagnetic phase, the magnetically charge ordered phase, and the $\sqrt{3}\times\sqrt{3}$ phase for $h/D<0.5$. These are shown as purple closed circles and blue triangles in Fig.~\ref{fig:phase_diag}.

\subsection{Stronger transverse field $0.5<h/D<0.65$ \label{sec:strong_h}}

\begin{figure*}
\includegraphics[width = \textwidth]{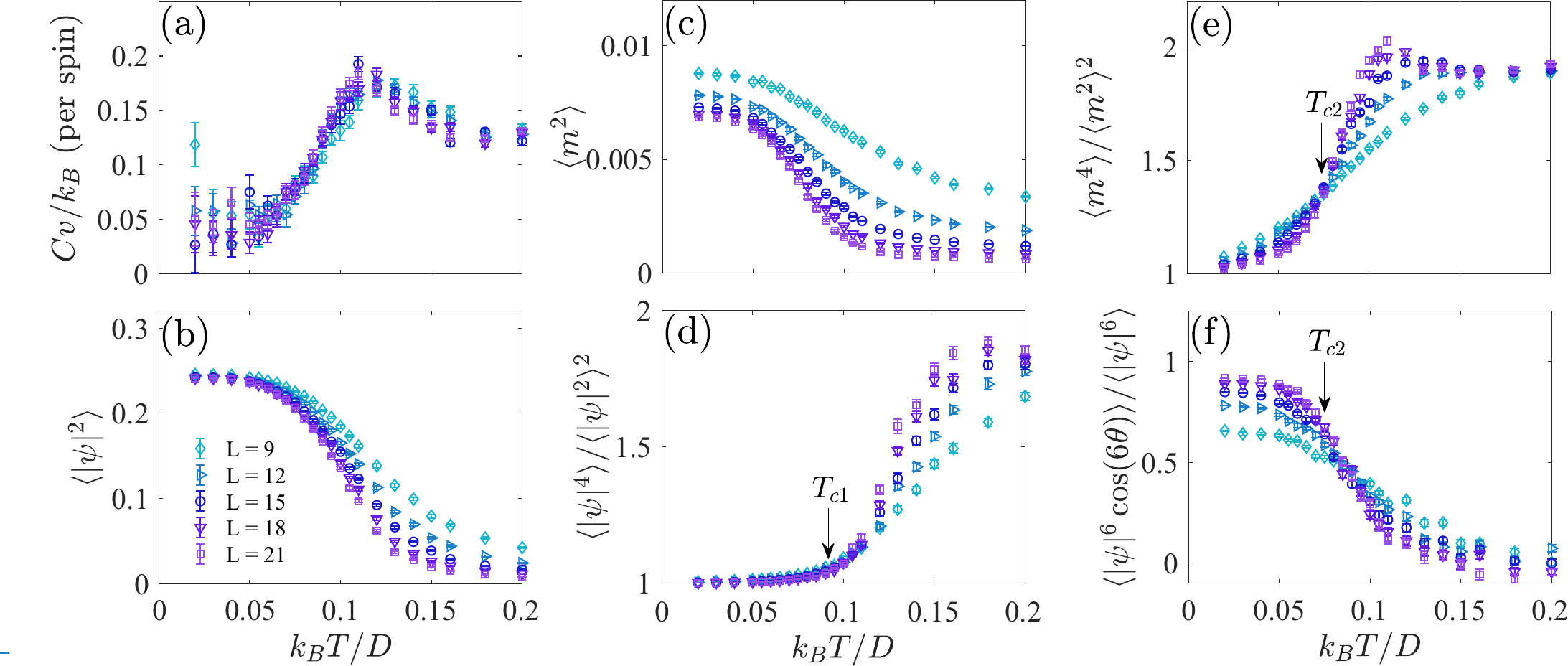}
\caption{Similar to Fig.~\ref{fig:h_0p25} but for transverse field $h/D = 0.6$. Arrows mark the critical temperatures estimated from data collapse. \label{fig:h_0p6}}
\end{figure*}

Fig.~\ref{fig:h_0p6} presents a temperature scan of the phase diagram at constant transverse field $h/D =0.6$. Both $\langle m^2\rangle$ and $\langle |\psi|^2\rangle$ steadily increases as $T$ decreases (Fig.~\ref{fig:h_0p6}b\&c). The anisotropy measure remains positive and approaches 1 as $T\to 0$ (Fig.~\ref{fig:h_0p6}f). These suggest that the system settles into the $\sqrt{3}\times\sqrt{3}$ phase at low temperature, which is similar to the case with $h/D = 0.25$. However, different from the previous case, the onset temperature of $\langle|\psi|^2\rangle$ is clearly higher than that of $\langle m^2\rangle$. 

The specific heat $C_v$ and the Binder ratio show further differences in comparison with the $h/D = 0.25$ data. $C_v$ does not show any sharp peaks except for a bump at the temperature scale $k_BT/D \approx 0.11$ (Fig.~\ref{fig:h_0p6}a). The bump at the lower temperature scale $k_BT/D \approx 0.05$ is most likely spurious due to our choice of the specific heat approximant Eq.~\eqref{eq:cv_approx}. The Binder ratio of $\psi$ for various system sizes $L$ do not show a crossing behavior typical for second order phase transitions; instead, their values approximately converge at a temperature scale $k_BT/D \approx 0.1$ (Fig.~\ref{fig:h_0p6}d). By contrast, the Binder ratio of $m$ for different system sizes $L$ shows a clear crossing behavior at slightly lower temperature (Fig.~\ref{fig:h_0p6}e). The anisotropy measure data cross at about the same temperature as the Binder ratio of $m$ although the limited data quality and the finite size effect make it difficult to pinpoint the crossing temperature (Fig.~\ref{fig:h_0p6}f).

Taken together, the data point to the alternative melting pathway II shown in Fig.~\ref{fig:landau_theory}b, i.e. the $\sqrt{3}\times\sqrt{3}$ magnetic order melts through a floating KT phase in close analogy with the six-state clock model~\cite{Jose1977,Challa1986}: First of all, the specific heat bump is reminiscent of the specific heat of the six-state clock model. Secondly, in the floating KT phase, the critical fluctuations in $\psi$ result in the ``converging" behavior of its Binder ratios near the transition from the paramagnetic phase to the floating KT phase. Thirdly, the transition from the floating KT phase to the $\sqrt{3}\times\sqrt{3}$ phase is associated with the long range ordering of $\theta$. The crossing of the anisotropy measure of $\psi$ reflects the ordering of $\theta$. Finally, since the order parameter $m$ is coupled to $\theta$ (Eq.~\eqref{eq:landau}), the ordering in $\theta$ induces the ordering of $m$. The crossing of the Binder ratio of $m$ therefore mirrors the crossing of the anisotropy measure.

\begin{figure*}
\includegraphics[width = 0.75\textwidth]{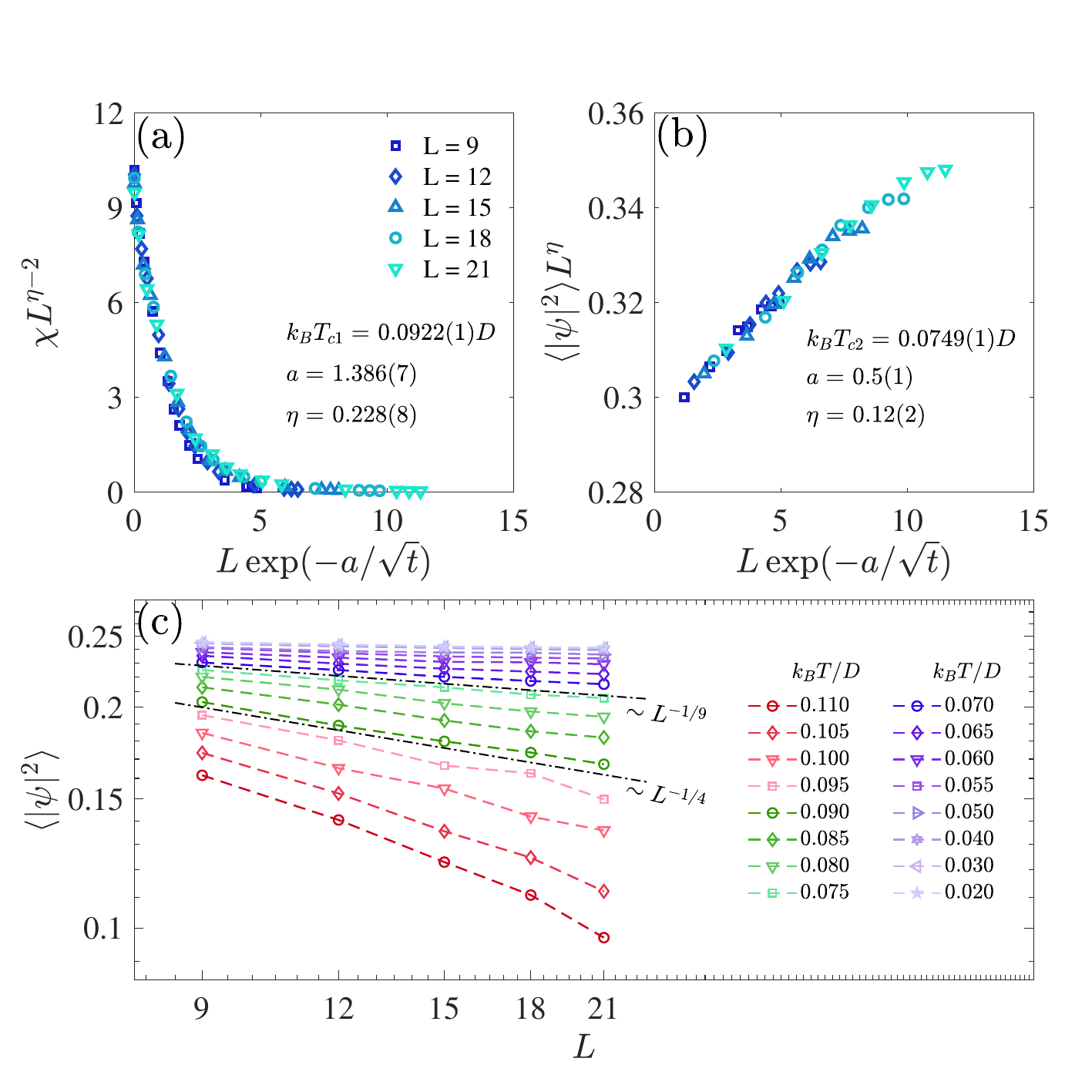}
\caption{(a) Data collapse for the effective susceptibility (see the main text for definition) $\chi$ above the high temperature phase transition $T_{c1}$ assuming the KT universality. Reduced temperature $t = (T-T_{c1})/T_{c1}$. (b) Data collapse for $\langle|\psi|^2\rangle$ below the low temperature phase transition $T_{c2}$ assuming the KT universality. Reduced temperature $t = (T_{c2}-T)/T_{c2}$. (c) Log-log plot of $\langle|\psi|^2\rangle$ as a function of system size at various temperatures. Data points fall into the paramagnetic phase, the floating KT phase, and the $\sqrt{3}\times\sqrt{3}$ phase are shaded in red, green, and purple, respectively. Dashed lines show the expected scaling at $T_{c1}$ ($L^{-1/4}$) and $T_{c2}$ ($L^{-1/9}$).
\label{fig:h_0p6_data_collapse}}
\end{figure*}

We test the validity of the above picture by performing data collapse for relevant observables. The transition from the paramagnetic phase to the floating KT phase, and from the floating KT phase to the $\sqrt{3}\times\sqrt{3}$ phase, are both expected to be in the KT universality class. We first consider the transition from the paramagnetic phase to the floating KT phase. The relevant observable is the effective susceptibility $\chi \overset{\mathrm{def}}{=} N\langle|\psi|^2\rangle/(k_BT)$ in the paramagnetic phase~\cite{Challa1986}. Fig.~\ref{fig:h_0p6_data_collapse}a shows our best attempt at the data collapse of $\chi$ assuming the KT universality. We find the transition temperature $T_{c1} \approx 0.0922$. The anomalous dimension $\eta \approx 0.228$, which is close to the expected value of $1/4$~\cite{Jose1977}. 

For the lower temperature transition from the floating KT phase to the $\sqrt{3}\times\sqrt{3}$ phase, the relevant observable is the magnitude of order parameter $\langle|\psi|^2\rangle$ in the $\sqrt{3}\times\sqrt{3}$ phase~\cite{Challa1986}. Fig.~\ref{fig:h_0p6_data_collapse}b shows our best attempt at the data collapse of $\langle|\psi|^2\rangle$ assuming the KT universality. The transition temperature $T_{c2} \approx 0.0749$. The anomalous dimension $\eta \approx 0.12$, which is close to the expected value of $1/9$~\cite{Jose1977}.

Having shown that the both transitions are consistent with the KT universality, we examine the correlation of $\psi$ in the floating KT phase. The log-log plot of $\langle|\psi|^2\rangle$ suggests that $\langle |\psi|^2\rangle$ seems to decay algebraically as a function of $L$ in the floating KT phase within the limited range of $L$~\cite{Challa1986}. In particular, the slope is close to $1/4$ near $T_{c1}$ and $1/9$ near $T_{c2}$, both in agreement with the expected anomalous dimension of $\psi$ at these transitions. 

To recapitulate, at the transverse field $h/D = 0.6$, our data suggest the system first enters the floating KT phase and then the $\sqrt{3}\times\sqrt{3}$ phase through two consecutive KT transitions, which is in agreement with the melting pathway II shown in Fig.~\ref{fig:landau_theory}b. 

At other values of $h/D>0.5$, we find similar behaviors as $h/D = 0.6$. We estimate the phase boundaries of the floating KT phase by performing data collapse assuming the KT universality class. The limited system size and quality of statistics do not allow us to put a stringent bound on $\eta$ in the data collapse. The low temperature transition at $T_{c2}$ proves to be particularly challenging due to the narrow temperature range between 0 and $T_{c2}$. The two KT transitions being close to each other poses further problems. We note similar issues arise in the QMC study of antiferromagnetic quantum Ising model on triangular lattice~\cite{Isakov2003}, where the authors resort to other means when estimating critical temperatures. Here, we use the theoretical value of $\eta = 1/4$ and $\eta = 1/9$ respectively for the higher and lower temperature KT transitions in the data collapse. The estimated $T_{c1}$ and $T_{c2}$ are shown as cyan lines in Fig.~\ref{fig:phase_diag}. 

\subsection{Intermediate transverse field $h/D=0.5$ \label{sec:interm_h}}

\begin{figure*}
\includegraphics[width = \textwidth]{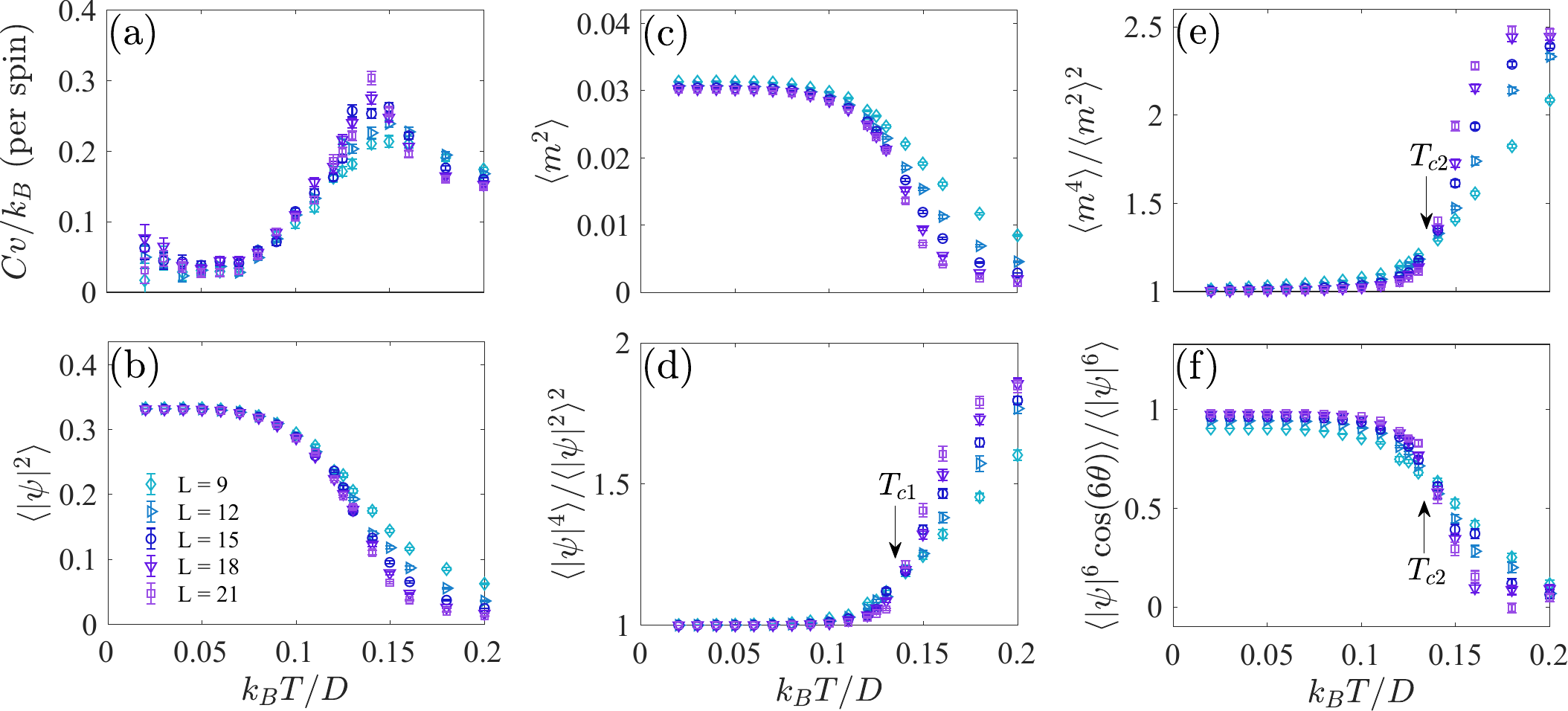}
\caption{Similar to Fig.~\ref{fig:h_0p25} but for transverse field $h/D = 0.5$. Arrows mark the critical temperatures estimated from the crossing points of the Binder ratio.
\label{fig:h_0p5}}
\end{figure*}

In the previous subsections, we have shown that the $\sqrt{3}\times\sqrt{3}$ phase melts through the intermediate magnetic charge ordered phase for small $h/D$, and through the floating KT phase for larger value of $h/D$. Crucially, the phase transitions are of different universalities. For small $h/D$, the high temperature transition is of the Ising universality, whereas the low temperature transition is believed to be in the three-state Potts universality class. For large $h/D$, both phase transitions are of KT universality. 

It is then natural to ask how the two distinct sequence of phase transitions are connected as we tune $h/D$. There are two possibilities. As we increase $h/D$ from 0, the Ising transition and the Potts transition move closer in temperature, and they eventually merge into a single first order phase transition. Upon further increasing $h/D$, the first order transition splits off into two KT transitions similar to a generalized six-state clock model~\cite{Cardy1980,Dorey1999}. Alternatively, the Ising transition,  the Potts transition, and the two KT transitions could all meet at a single multicritical point as suggested in Ref.~\onlinecite{Damle2015} --- the said multicritical point results from the interplay between the two intertwined order parameters $m$ and $\psi$ and is believed to be absent in the generalized six-state clock model.

We explore these aspects by scanning the phase diagram at $h/D = 0.5$ (Fig.~\ref{fig:h_0p5}). The specific heat $C_v$ shows a single peak that grows with the system size (Fig.~\ref{fig:h_0p5}a), which is typical for a second order phase transition. The onset of $\langle m^2\rangle$ and $\langle|\psi|^2\rangle$ seem to occur at about the same temperature (Fig.~\ref{fig:h_0p5}b\&{}c). We estimate the critical temperature associated with the ordering of $\psi$ and $m$ by using the crossing point of their respective Binder ratio. We find $k_BT_{c1}/D\approx 0.133(5)$ and $k_BT_{c2}/D\approx 0.136(5)$, which are identical within error bars. 

Our data suggest that the temperature scan at $h/D = 0.5$ must pass closely by the multicritical point, or the first order transition. If there is indeed a first order phase transition, it should occupy a relatively narrow window on the $h/D$ axis, and is unlikely to be strongly first order. We cannot make any further statements due to the limited system size and the long autocorrelation time of the QMC algorithm at this value of $h/D$.

\section{Discussion \label{sec:discussion}}

To conclude, our QMC simulation of Eq.~\eqref{eq:hamiltonian} shows that we may tune the two-step melting process of the $\sqrt{3}\times\sqrt{3}$ phase by quantum fluctuations. At the weak transverse field, the $\sqrt{3}\times\sqrt{3}$ phase melts through the intermediate magnetic charge order. This process is connected to the classical limit~\cite{Moeller2009,Chern2011,Chern2012}. By contrast, at relatively large transverse field, a distinct melting process emerges --- the $\sqrt{3}\times\sqrt{3}$ phase melts through the floating KT phase by two successive KT transitions.

Our work thus reveals the interesting prospect of controlling the thermal melting processes of complex magnetic orders through quantum fluctuations. Yet, a couple of important questions remain unanswered. The behavior of the model Eq.~\eqref{eq:hamiltonian} near the intermediate transverse field $h/D = 0.5$ is unclear. Specifically, it is unknown if the two aforementioned melting processes in this model are separated by a first order phase transition or a multicritical point. In addition, the zero temperature quantum phase transition from the $\sqrt{3}\times\sqrt{3}$ phase to the paramagnetic phase is not determined in this work. Given that accessing the moderate to large transverse field regime is challenging for Trotter-type QMC algorithms, stochastic series expansion algorithms that are tailored for models with long range interactions \cite{Sandvik2003} or with geometrical frustration \cite{Biswas2016,Biswas2018} may prove useful in tackling these questions. One could also reduce the time cost for evaluating the interaction energies by employing a clocked factorized Metropolis filter~\cite{Michel2019}.

Given that the Ising models with short range interactions are more amenable to Monte Carlo simulations, it would be interesting to explore the melting processes of the $\sqrt{3}\times\sqrt{3}$ phase there as well. Extensive Monte Carlo simulation of the classical Ising model on triangular and kagome lattices with first, second, and third neighbor interactions have clarified how the two-step melting processes merge into a single first order melting transition~\cite{Damle2019}. It is then natural to examine the impact of quantum fluctuations on these systems~\cite{Biswas2018} with an eye toward the multicritical point~\cite{Damle2015}.

In light of the Ho$^{3+}$ based tripod kagome magnet~\cite{Dun2019}, our work suggests that one may potentially explore these distinct melting sequences in a thin film of Mg$_2$Ho$_3$Sb$_3$O$_{14}$ or similar systems by tuning the relative strength of the dipole interaction energy scale $D$ with respect to the quantum fluctuation energy scale $h$. This may be achieved by epitaxial strain from the substrates~\cite{Middey2016} or by chemical pressure.

From the material perspective, while the minimal model Eq.~\eqref{eq:hamiltonian} captures the competition between the magnetic dipole interaction and the quantum fluctuations, a few important features of the tripod kagome magnets are not accounted for. The spin quantization axis $\hat{z}_i$ cant away from the kagome plane in the tripod kagome materials by about $26^\circ$~\cite{Dun2016,Dun2017,Paddison2016}. As the magnetic dipole interaction depends on the configurations of $\hat{z}_i$, one would expect the canting angle to have significant impact on the physics of Eq.~\eqref{eq:hamiltonian}. Indeed, the classical limit of the model Eq.~\eqref{eq:hamiltonian} is known to exhibit dramatically different physics when the canting angle varies. On one hand, when $\hat{z}_i$ are in the kagome plane, the model hosts the $\sqrt{3}\times\sqrt{3}$ magnetic ground state~\cite{Moeller2009,Chern2011,Chern2012}. On the other hand, when $\hat{z}_i$ are perpendicular to the kagome plane, the system is very glassy and shows different magnetic long range orders (the so-called ``Figure Seven" state)~\cite{Chioar2014,Chioar2016,Hamp2018}. The tripod kagome material interpolates these two limits, and it is not clear \emph{a priori} which limit is closer to the material reality. The experiment on Mg$_2$Dy$_3$Sb$_3$O$_{14}$ suggests that the former in-plane limit is perhaps a more appropriate starting point for theoretical discussions~\cite{Paddison2016}.

Another important feature of the tripod kagome materials is their three dimensionality. Since the floating KT phase is absent in three dimensions, one may expect a direct continuous phase transition from the $\sqrt{3}\times\sqrt{3}$ phase to the paramagnetic phase in the 3D XY universality class. Alternatively, the system may develop long range order in the kagome plane but remain short range correlated between the planes. In addition, the subtle interplay between the canting and the three dimensionality plays a crucial role in understanding the tripod kagome magnets. In the classical tripod kagome magnet Mg$_2$Dy$_3$Sb$_3$O$_{14}$, the interlayer coupling stabilizes the magnetic charge order and helps the system evade the spin glass physics that would otherwise occur in systems with large canting angles~\cite{Paddison2016,Hamp2017}.

Finally, the Ho$^{3+}$ ion carries large nuclear spin and significant hyperfine coupling~\cite{Krusius1969}. It has been argued that the hyperfine coupling is responsible for the magnetic ordering in Ho$^{3+}$ based garnet Ho$_3$Ga$_5$O$_{12}$~\cite{Paddison2019} and suppresses the quantum coherence in tripod kagome magnet Mg$_2$Ho$_3$Sb$_3$O$_{14}$~\cite{Dun2019}. The role of the hyperfine coupling thus requires careful theoretical assessment as well.

To conclude, we expect exploring these aspects will reveal more interesting physical effects in tripod kagome magnets and pave the way for a deeper understanding of this material family.

\begin{acknowledgments}
Yuan Wan thanks Martin Mourigal and Joseph Paddison for informing him their experiments, and Zi-Yang Meng for collaborations at the early stage of this work. We are grateful to Kedar Damle and Claudio Castelnovo for illuminating discussions and a critical reading of the manuscript.  The QMC simulations were carried out on TianHe-1A at the National Supercomputer Center in Tianjin, China. This work is supported in part by the National Science Foundation of China (Grant No. 11974396, Wan) and the International Young Scientist Fellowship from the Institute of Physics, Chinese Academy of Sciences (Grant No. 2018004, Humeniuk).
\end{acknowledgments}

\bibliography{quantum_kagome}

\end{document}